\documentclass[aps,prb,floats,twocolumn,showpacs,amsmath,amsfonts,amssymb]{revtex4}
\usepackage{graphicx}
\usepackage{dcolumn}
\usepackage{bm}
\begin{document}
\title{Spin effects in transport through non-Fermi liquid quantum dots}
\author{Fabio Cavaliere$^{1,3}$ Alessandro Braggio$^{2}$, Maura
  Sassetti$^{1}$, and Bernhard Kramer$^{3}$ \vspace{1mm}}
\affiliation{$^{1}$Dipartimento di Fisica Universit\`{a}
  di Genova, INFM-LAMIA, Via Dodecaneso 33, I-16146 Genova\\ $^{2}$ Institut
  fur Theoretische Physik III, Universitatsstrasse 150
  D-44801 Bochum \\ $^{3}$ I. Institut f\"ur
  Theoretische Physik, Universit\"at Hamburg,
  Jungiusstra\ss{}e 9, D-20355 Hamburg\\
  \vspace{3mm}} \date{May 18, 2004}
\begin{abstract}
  The current-voltage characteristic of a one dimensional quantum dot
  connected via tunnel barriers to interacting leads is calculated in the
  region of sequential tunneling. The spin of the electrons is taken into
  account.  Non-Fermi liquid correlations implying spin-charge separation are
  assumed to be present in the dot and in the leads. It is found that the
  energetic distance of the peaks in the linear conductance shows a
  spin-induced parity effect at zero temperature $T$. The temperature
  dependence of the positions of the peaks depends on the non-Fermi liquid
  nature of the system. For non-symmetric tunnel barriers negative
  differential conductances are predicted, which are related to the
  participation in the transport of collective states in the quantum dot with
  larger spins. Without spin-charge separation the negative differential
  conductances do not occur. Taking into account spin relaxation destroys the
  spin-induced conductance features. The possibility of observing in
  experiment the predicted effects are briefly discussed.
\end{abstract}
\maketitle 

\section{Introduction}
\label{sec:1}

Spin phenomena in the transport properties of low-dimensional quantum systems
have become a subject of increasing interest.\cite{p98,a98} Several important
effects have been found when controlling transport of electrons one by one in
quantum dots. For instance, a parity effect in the Coulomb blockade\cite{t96}
of quantum dots with very small numbers of electrons, and in Carbon
nanotubes\cite{cetal98,betal02} have been detected.  The spin blockade effect,
especially in the non-linear current-voltage  characteristic of
one-dimensional (1D) quantum dots has been predicted.\cite{wetal94,w95}
Combining the spin blockade with spin-polarized detection, the electron spin
in a lateral quantum dot has been probed, and spin-related phases were found
that have been associated with correlations between the
electrons.\cite{cetal02}

The former parity effect is related to the Pauli principle and is
quantitatively affected by the contribution of the exchange interaction
towards the energy of the ground state of the electrons occupying the quantum
dot. The spin blockade effect, which leads to a negative differential
conductance (NDC) in the current as a function of the bias voltage, is due to
a combined influence of the exchange contribution to the energies of the
correlated electronic eigenstates and the spin selection rules for the
transport processes. For example, an excited $n$-electron state of a quantum
dot generally can be depopulated via two spin channels, namely either by
increasing or by decreasing the $z$-component of the total spin, $s$, by
$1/2$. However, when the state {\em with the highest total spin} becomes
populated at a certain bias voltage $V$, it can be depopulated only via
processes that decrease $s$. This can lead to a reduction of the
total current $I(V)$, when increasing the bias voltage, thus
yielding ${\partial}I/{\partial}V < 0$.

This phenomenon has been predicted by using a simple model for the
transport mediated by sequential electron tunneling processes through a 1D
quantum dot containing few electrons ($n\leq 4$). The electronic
eigenstates of the latter have been determined numerically exactly in the
presence of interactions and spin.\cite{wetal94,w95} The tunneling matrix
elements of the barriers connecting the states of the quantum dot to those in
the Fermi-liquid leads were assumed to include spin selection rules via
Clebsch-Gordan coefficients and assuming {\em ad hoc} some tunneling coupling
matrix elements between the correlated quantum dot states and the states in
the leads. Only later, the matrix elements have been calculated
microscopically in order to test the former assumptions.\cite{jauregui} There
are several, in the details somewhat different and intricate mechanisms that
produce such a spin-induced reduction of the current. Characteristic
dependences on the magnetic field may be used to distinguish between them.
While certain signatures of these features have been found in a recent
experiment done on 2D quantum dot,\cite{blick} experimentally well-controlled
evidence in quasi-1D quantum dots is missing.

The recent experimental realization of semiconductor-based 1D quantum
wires\cite{ths95,yetal96,yetal97} has opened new perspectives to
systematically investigating the influence of interactions, spins and
impurities on electron transport
properties. Signatures of spin-charge separation have
been observed \cite{parwi} and analyzed \cite{tserk} in the tunneling
between parallel one-dimensional wires. Also carbon nanotubes can now be
controlled to such a high degree that investigations of electronic transport
features have become possible.\cite{cetal98,betal02,dek01,lbp02}
The effects of
non-Fermi liquid correlations in 1D quantum dots have been analyzed
non perturbatively in the {\em coherent} tunneling
regime.\cite{yuli,poly}

In the non-linear transport spectra of these devices, obtained as the
derivative of the current-voltage characteristics at different gate voltages,
a large number of low-energy excited states have been found.\cite{a00} These
cannot be understood only in terms of charge excitations\cite{k00} but are
theoretically predicted to be related also to the spin.\cite{b01} Thus, one
can expect that 1D spin blockade effects could be seen in an experiment. Among
the experimental realizations of 1D quantum dots are electron islands between
two successive impurities in a 1D quantum wire (containing interacting
electrons) and carbon nanotubes. Thus an extension to the theory of the
blockade effects should include not only the generalization to higher electron
densities---the previous calculations have been done in the limit of low
electron density---but treat the interactions and the spins within the quantum
dot and within the leads on an equal footing.

There are perspectives for application of quantum structures in
spin-electronics, quantum computing and communication.\cite{i99} Previous
works focused on spin transport in (2D) quantum dots connected to
non-interacting leads in the presence of a magnetic field,\cite{r00} including
an oscillating magnetic electron spin resonance component.\cite{en01} Spin
transport in circuits with ferromagnetic elements and Luttinger-liquid
interaction\cite{br00,s98,b00} has been considered. In view of the
applications, the theory of the spin control of electron transport in the
presence of correlations is very important since in nanoscale devices the
latter can be very important. In our previous work,\cite{b01} we have found
that at $T=0$, spin polarization effects are robust against the
correlations. They can even be enhanced, if the polarization is not complete.

In the present paper, we extend our previous calculations to include
temperature effects, asymmetry in the tunnel barriers and the
effect of spin relaxation. The former are treated microscopically within the
sequential tunneling approximation using the microscopic model of a 1D quantum
dot described by a Luttinger liquid of finite length connected via
tunnel barriers to semi-infinite interacting leads. Spin
relaxation is treated phenomenologically by including tunnel rates
corresponding to spin-flip transitions into the master equation for the
dynamics of the probability distribution. Our main results are
\begin{enumerate}
\item We deduce that the temperature dependence of the linear conductance
  peaks reflects the non-Fermi liquid correlation in the leads.
\item In non-linear transport, we find that many-particle states with higher
  spins in the quantum dot act as traps such that the electric current
  decreases with increasing bias voltage. The physical origin of this is
  spin-charge separation.
\item Spin relaxation leads to a destruction of the negative differential
  conductance peaks. This confirms that they are related to the trap
  properties of higher-spin states.  A typical ``phase diagram'' of the
  crossovers between negative and positive differential conductance peaks is
  calculated.
\end{enumerate}  
We discuss the quantitative conditions for observing the predicted effects in
an experiment.

In the next section, we introduce the model for the Luttinger liquid quantum
dot. Section 3 is used to discuss the energy scales of the system. In the
section 4 we describe the approach for calculating the transport properties.
Section 5 is devoted to the linear and section 6 to the non-linear regime. In
section 7 we compare with other recent approaches, discuss possible
experimental realizations and draw some conclusions.

\section{Tomonaga-Luttinger model for a 1D quantum dot with spin}
\label{microscopicmodel}

\begin{figure}[htbp]
\setlength{\unitlength}{1cm}
\includegraphics[width=7cm]{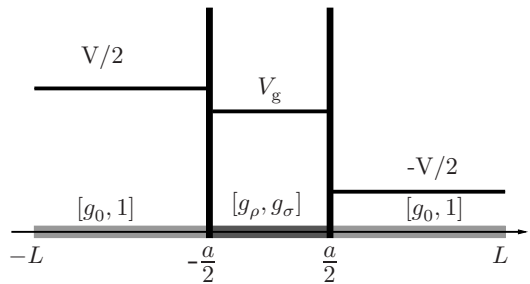}
\par
    \centering
\caption{Schematic representation of the Luttinger liquid quantum dot
  ($|x|<a/2$) connected via tunneling barriers to semi-infinite wires. The
  charge ($g_{\rho}$) and spin ($g_{\sigma}$) interaction parameters in the
  leads and the dot are assumed to be different. The bias voltage $V$
  controls the chemical potential difference of the leads.  The gate voltage
  $V_{\rm g}^{}$ controls the chemical potential quantum dot (cf. Fig.
  \ref{fig:circuit}).}
  \label{fig:dotmodel}
\end{figure}

We consider the 1D system shown schematically in Fig.~\ref{fig:dotmodel}. The
region $|x| < a/2$ represents a finite quantum wire that plays the role of the
1D quantum dot. This is assumed to be connected to reservoirs with
electrochemical potentials $\mu_{\lambda}=\pm eV/2$ ($\lambda= \rm R, \rm L$)
via semi-infinite 1D systems between $-L < x < -a/2$ and $a/2 < x < L$
($L\to\infty$) that represent left (L) and right (R) leads, respectively. The
electrochemical potentials are controlled by the bias voltage $V$.
Tunneling barriers at $x=\pm a/2$ connect the quantum dot with the leads.
The three regions of the quantum wire are assumed as interacting Luttinger
liquids (LL)\cite{t50,l63,h81,v95} with possibly {\em different} interaction
constants for charges ($g_{\rho}$) and spins ($g_{\sigma}$)
(Fig.~\ref{fig:dotmodel}).\cite{ss9y,ketal02} In addition to the bias
voltage $V$ we assume that the chemical potential in the dot region is tuned
via a gate voltage $V_{\rm g}$. The total Hamiltonian is
\begin{equation}
\label{eq:TheHamiltonian}
H=H_{0}+H_{\rm t}+H_{\rm c}\, .
\end{equation}
Here, $H_{0}^{}=H_{0}^{\rm(d)}+H_{0}^{\rm(L)}+H_{0}^{\rm(R)}$ describes the
three uncoupled LL, $H_{\rm t}$ the tunneling term, and $H_{\rm c}$ the
coupling with the external electric circuit.

The 1D quantum dot ($|x|<a/2$) is treated with open boundary
conditions,\cite{fabrizio,eggert} $\psi_{s}^{\rm(d)}(\pm a/2)=0$, with the
fermion operators $\psi_{s}^{\rm(d)}(x)$ corresponding to the $z$-component of
the spin $s=\pm 1$ (unit $\hbar/2$).

The Luttinger liquid theory describes the energetically
low-lying excitations of 1D interacting electrons around the Fermi
energy $E_{\rm F}$. It relies on the linearization of the dispersion
relation around the Fermi level giving rise to a bare, constant level
spacing $\varepsilon_0=\hbar\pi v_{\rm F}/a$. As a necessary condition
for this assumption, one needs to fulfill the condition $E_{\rm
F}\gg\varepsilon_0$. For a system of finite length, this implies a
constraint between the Fermi wave number and the momentum
discretization, $k_{\rm F}\gg \pi/a$. There is also a constraint with
respect to temperature induced by the finite size of the dot. In order
to resolve the energy levels of the correlated states the temperature
should be lower than the level spacing, $k_{\rm B}T\ll \varepsilon_{0}$.

Near the Fermi point $k_{\rm F}$ the fermion operators are decomposed into
fields that propagate to the right (${\rm r}$) and to the left (${\rm l}$)
\begin{equation}
\psi_{s}^{\rm(d)}(x)=e^{ik_{\rm F}x}\psi_{s,\rm r}^{\rm(d)}(x)+
e^{-ik_{\rm F}x}\psi_{s,\rm l}^{\rm(d)}(x)\,.
\end{equation}
Because of the boundary conditions these fields are not independent. The
bosonization description is done using only $\psi_{s,\rm r}^{\rm(d)}(x)$. The
system is diagonalized in the presence of interactions. The corresponding
Hamiltonian is\cite{fabrizio,eggert} (units such that $\hbar = 1$)
\begin{equation}
\label{eq:dotfreehamil} 
H_{0}^{\rm(d)}=\sum_{\nu=\rho,\sigma}\sum_{q>0}\omega_{\nu}(q)
\nu^{\dagger}(q)\nu(q)
+ \frac{\pi}{4a}\left[\frac{v_{\rho}}{g_{\rho}} {\hat n}^{2} + 
\frac{v_{\sigma}}{g_{\sigma}} {\hat s}^{2}\right] \, . 
\end{equation}
Here, $\nu^{\dagger}(q)$, $\nu^{}(q)$ are the boson operators of the
collective charge ($\nu = \rho$) and spin ($\nu = \sigma$) density waves (CDW
and SDW). Due to the boundary condition the wave number is quantized, $q=\pi
m/a$ ($m$ integer $\ge 1$). The energy spectra are\cite{v95}
\begin{equation}
\omega_{\nu}^{}(q) = v_{\nu}^{}q\,,\qquad v_{\nu}^{}=\frac{v_{\rm F}}{g_{\nu}}
\left(1+V_{\rm ex}^{}\right)
\end{equation}
with the interaction parameters
\begin{equation}
\label{eq:grho}
g_{\rho}^{2} = \frac{1 + V_{\rm ex}}{1 - V_{\rm ex} + 4 V_{0}} \,,\quad
g_{\sigma}^{2} = \frac{1 + V_{\rm ex}}{1 - V_{\rm ex}}
\end{equation}
where
\begin{equation}
\label{eq:luttg1}
V_{0}^{}=\frac{\hat{V}(0)}{2\pi v_{\rm F}}\,,\qquad 
V_{\rm ex}^{}=\frac{\hat{V}(2k_{\rm F})}{2\pi v_{\rm F}}\, 
\end{equation}
are proportional to the forward ($q\to 0$), and part of the backward ($q\to
2k_{\rm F}$) contributions of the electron interaction, respectively. The
quantity $\hat{V}(q)$ denotes the Fourier transform of the interaction
potential inside the dot. The backward term corresponds to an exchange
interaction. The parameters fulfill $V_{\rm ex} < 1$, necessary in order to
have a bounded Hamiltonian, and $V_{0} > V_{\rm ex}$. This implies $0 <
g_{\rho} \le 1$ (repulsive charge-charge interaction) and $g_{\sigma} \ge 1$.

The zero mode operators ${\hat n}$ and ${\hat s}$ represent the excess
number of charges, and of the $z$-component of the total spin with respect to
their average values on the ground state. The latter correspond to
$n_0=2k_{\rm F}a/\pi-1$ and $s_0=0$ in the absence of a magnetic field. The
eigenvalues of the zero-mode operators are integers $n,s$ with the constraint,
due to the boundary conditions, $n+s = {\rm even}$. The zero-mode energy
contributions in (\ref{eq:dotfreehamil}) represent the energy needed for
changing the total charge and spin with respect to the ground state.

For the leads we assume a LL with open boundary conditions at the tunnel
barriers. At $L=\pm\infty$ they are assumed to be connected to reservoirs with
different electrochemical potentials. The difference of the latter is
proportional to the bias voltage.
In state-of-the-art semiconductor quantum wires (and also in carbon nanotube
systems) the exchange is only a very small correction as compared to the
Coulomb part of the interaction\cite{sk98} which in any case is expected to be
considerably weaker that in the region of the quantum quantum dot. The
exchange interaction in the leads is not expected to influence the results
described below on the qualitative level. Especially, it is not expected to
influence the positions of the conductance peaks that are determined by the
interactions in the region of the quantum dot. At most, it can give a small
quantitative correction of the dependence on temperature of the conductance
peaks which is determined by the global properties of the system as has been
discussed earlier.\cite{ketal02} Therefore, we neglect in this paper the
exchange interaction in the leads completely for the sake of simplicity.
The Hamiltonian describing the excitations in the leads is then\cite{v95}
\begin{equation*}
\begin{split}
  H_{0}^{(\lambda)}=
\sum_{\nu=\rho,\sigma}^{}&\sum_{k>0}\Omega_{\nu}(k)\, 
\nu_{\lambda}^{\dagger}(k) \nu_{\lambda}^{}(k)\\
  &+\frac{\pi {\bar v}_{\rm
      F}}{4L}\left[\frac{1}{g_{0}^{2}}\,
{\hat n}_{\lambda}^{2}+{\hat s}_{\lambda}^{2}\right]\, ,
\end{split}
\end{equation*}
with ${\bar v}_{\rm F}$ the Fermi velocity in the leads,
$\nu_{\lambda}^{\dagger}(k)$, $\nu_{\lambda}^{}(k)$, the boson operators of
collective CDW, SDW and ${\hat n}_{\lambda}$, ${\hat s}_{\lambda}$ the zero
mode operators for the excess of charge and spin with respect to their average
values in the leads. The energy spectra are ($k=\pi m /L$, $m$ integer $\geq
1$)
\begin{equation*}
\Omega_{\rho}^{}(k)=\frac{{\bar v}_{\rm F}}{g_{0}^{}}k\,
,\qquad\Omega_{\sigma}^{}(k)={\bar v}_{\rm F}^{}k\,
\end{equation*}
with the charge interaction parameter $g_{0}$ 
\begin{equation*}
g_{0} = 
\left({1+\frac{2U_{0}^{}}{\pi {\bar v}_{\rm F}}}\right)^{-1/2}<1
\end{equation*}
determined by the average interaction $U_{0}$ in the leads.

The coupling between the leads and the dot is described by tunnel barriers at
$x_{\rm L}=-a/2$ and $x_{\rm R}=a/2$
\begin{equation}
\label{eq:tunham}
H_{\rm t}= \sum_{s=\pm 1}\sum_{\lambda={\rm L,R}} 
\left[t_{\lambda}^{} \psi_{s,\rm r}^{{\rm(\lambda)}
  \dagger}(x^{}_{\lambda})\psi_{s,\rm r}^{\rm(d)}(x^{}_{\lambda})+ {\rm
  h.c.}\right]\,,
\end{equation} 
with the right moving fermion operators normalized to the shortest
wavelength, and $t_{\rm L,R}$ the transmission amplitudes of the
barriers.\cite{fabrizio}

\begin{figure}[htbp]
\setlength{\unitlength}{1cm}
\begin{center}
\includegraphics[width=6cm,keepaspectratio]{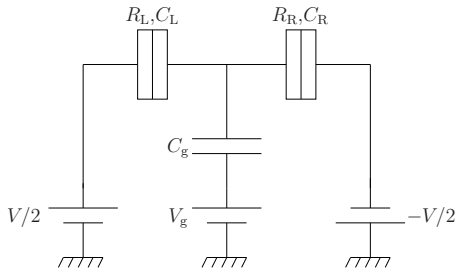}
\end{center}
\par
    \centering
\caption{Equivalent circuit for the quantum dot with asymmetric tunnel
  barriers. Left and right tunnel junctions are parameterized by the
  capacitances $C_{\rm L,R}$ and the resistances $R_{\rm L,R}\equiv
  \omega_{\rm c}^2/\pi e^2 t^2_{\rm L,R}$ (cf.~(\ref{eq:ratesum})),
  $\omega_{\rm c}$ cutoff energy of the leads, $t_{\rm L,R}$ transmission
  amplitudes of the barriers, $C_{\rm g}$ gate capacitance, $V_{\rm g}$ gate
  voltage, $V$ bias voltage.}
\label{fig:circuit}
\end{figure}

The operator of the external bias and gate voltages that allows to
electrically control current transport and charge density in the dot are
written in terms of the operators $\hat{n}$ of the excess charges
(Fig.~\ref{fig:circuit})\cite{nazarov}
\begin{equation}
\label{eq:hamem}
H_{\rm c}=\frac{eV}{2}\left[{\hat n}_{\rm R}-{\hat n}_{\rm
    L}\right]-e\left[\frac{\delta C}{2 C_{\Sigma}^{}}V + \frac{C_{\rm
    g}^{}}{C_{\Sigma}^{}}V_{\rm g}^{}\right]{\hat n}\,.
\end{equation}
Here, $V$ and $V_{\rm g}$ are the bias and the gate voltage,
$C_{\Sigma}=C_{\rm L}+C_{\rm R}+C_{\rm g}$ is the total capacitance with
$C_{\rm L}$,$C_{\rm R}$ and $C_{\rm g}$ the capacitances of the leads and the
gate, where $\delta C = C_{\rm L}-C_{\rm R}$.

\section{Energy scales}
\label{energyscales}

\begin{figure}[htbp]
\setlength{\unitlength}{1cm}
\begin{center}
\includegraphics[width=7cm,keepaspectratio]{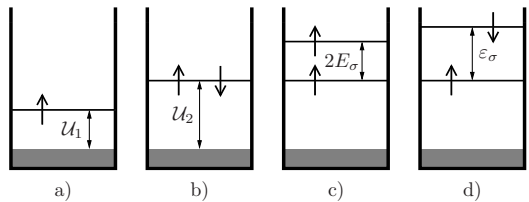}
\end{center}
\vspace{-0.5cm}
\caption{Schematic examples of excited 
  states with respect to a background charge $n_{0} = (2ak_{\rm F}^{}/\pi)-1$
  and spin $s_{0} = 0$. (a) $n=1$ and $s=1$, ${\mathcal U}_1\equiv {\mathcal
    U}(1,1,0,0)$ (see Eq. (\ref{eq:confen}), (b) $n=2$ and $s=0$, ${\mathcal
    U}_2\equiv {\mathcal U}(2,0,0,0)$, (c) excited state with $s=2$ for $n=2$,
  (d) spin excitation with respect to the state (b).}
\label{fig:states}
\end{figure}

The states of the isolated quantum dot are
$\left|n,s,\left\{l_{q}^{\rho}\right\},\left\{l_{q}^{\sigma}\right\}\right>$
where $n$ and $s$ are the excess numbers of charges and spins with respect to
the ground state with charge $n_0$ and spin $s_0$, and
$\left\{l_{q}^{\nu}\right\}$ the occupation numbers of the CDW ($\nu = \rho$)
and SDW ($\nu = \sigma$) at different $q$. Some examples of these states are
schematically shown in Fig.~\ref{fig:states}.
Figure \ref{fig:states}a represents the state $\left|1,1,\{0\},\{0\}\right>$,
corresponding to one excess charge $n=1$ with spin $s=1$. Figure
\ref{fig:states}b shows the state $\left|2,0,\{0\},\{0\}\right>$ with two
extra electrons and $s=0$. A higher spin state ($s=2$) is shown in
Fig.~\ref{fig:states}c.
Excited states with unchanged $n$ and $s$ contain CDW and/or SDW. Figure
\ref{fig:states}d shows an example with the smallest possible wave-number $q_0
= \pi / a$ and smallest excitation energy for $n = 2$ and $s=0$. The operator
\begin{equation*}
\sigma^{\dagger}\left(q_0\right)\left|2,0,\{0\},\{0\}\right>=
\left|2,0,\left\{0\ldots\right\},\left\{1,0,\ldots,0,\ldots\right\}\right>\,.
\end{equation*}
creates a linear superposition of the state (cf. Fig.~(\ref{fig:states}d)) and
the one with inverted spins. In the presence of interaction a charge
excitation with the smallest wave number would be associated with a much
larger excitation energy. Excitations with higher wave numbers induced by
operators $\nu^{\dagger}(q > q_0)$, as well as multiple excitations
$[\nu^{\dagger}(q)]^{r}$ ($r > 1$) are also possible.
  
Using the Hamiltonian (\ref{eq:dotfreehamil}) we write the total energy of the
particular configuration
$\left|n,s,\left\{l_{q}^{\rho}\right\},\left\{l_{q}^{\sigma}\right\}\right>$
as
\begin{equation}
\label{eq:confen}
{\mathcal U}\left(n,s,l_{\rho}^{},l_{\sigma}^{}\right) =
\frac{E_{\rho}^{}}{2} n_{}^{2} + \frac{E_{\sigma}^{}}{2} s_{}^{2} + 
l_{\rho}^{} \varepsilon_{\rho} + l_{\sigma}^{} \varepsilon_{\sigma}\,.
\end{equation} 
The first two terms represent the contributions of charge and spin additions.
The second two terms correspond to the CDW and SDW.  Since the excitation
spectra are linear in the wave number, the energies of the collective modes
depend on the total numbers of excitation quanta $l_{\nu}=\sum_{q}q
l_{q}^{\nu}$ only via the discrete excitation energies
$\varepsilon_{\nu}^{}=\pi v_{\nu}^{}/a$.  From the microscopic theory one gets
($\nu=\rho,\sigma$)
\begin{equation}
\label{charging}
E_{\nu}=\frac{\pi v_{\nu}}{2ag_{\nu}} = \frac{\varepsilon_0}{2}
\frac{1+V_{\rm ex}^{}}{g_{\nu}^{2}}\, ,
\end{equation}
where $\varepsilon_0=\pi v_{\rm F}/a$ the constant level spacing in the
non-interacting case. These energies are different from zero even without
interaction due to the discrete nature of the energy levels inside the dot and
the Pauli principle.

Despite the microscopic model provides quantitative estimates for the charge
and spin addition energies, several influences that occur in experimental
setups are here neglected. For instance, the coupling with the gates, long
range interaction effects, and the screening due to the nearby 2DEG present in
cleaved edge overgrowth systems\cite{a00,yetal97} affect
$E_{\rho}$ and can cause strong deviations from the simple expression
(\ref{charging}).~\cite{k00} More robust is the spin addition energy which is
influenced at most by the exchange part of the interaction. Therefore, we
treat in the following $E_{\rho}^{}$ as a free parameter with $E_{\rho} \gg
E_{\sigma}$.

Phenomenologically, one identifies $E_{\rho}$ with the total electrostatic
energy (Fig.~\ref{fig:circuit}) $E_{\rho} = e^2/C_{\Sigma}$ of the circuit
model. With this one includes the term $e n V_{\rm g}C_{\rm
g}/C_{\Sigma}$ coming from (\ref{eq:hamem}) into the zero mode of the charge sector in (\ref{eq:confen}),
$E_{\rho}{\left(n-n_{\rm g}^{}\right)}_{}^{2}/2$ with $n_{\rm g}=V_{\rm
  g}C_{\rm g}/e$.

For the spin addition energy we continue using the microscopic expression
(\ref{charging})
\begin{equation*}
E_{\sigma}=\frac{\varepsilon_0}{2}\left(1-V_{\rm ex}\right)\,.
\end{equation*}
Due to the exchange interaction this energy is decreased as compared to the
non-interacting case.  The energy quanta of the collective excitations can be
expressed microscopically as
\begin{equation}
\label{excitations}
\varepsilon_{\nu}=
\varepsilon_0 \frac{1 + V_{\rm ex}^{}}{g_{\nu}}\, .
\end{equation}

The presence of the exchange interaction renormalizes simultaneously the
energy $E_{\sigma}$ of the total spin configuration and the energy
$\varepsilon_{\sigma}$ of the spin waves
\begin{equation*}
\varepsilon_{\sigma}=\varepsilon_0{\sqrt{1-V_{\rm ex}^2}}=
2E_{\sigma}\sqrt{\frac{1+V_{\rm ex}}{1-V_{\rm ex}}}
\end{equation*}
The plasmon energy
$\varepsilon_{\rho}$ is, on the other hand, affected by the Coulomb
interaction with in general $\varepsilon_{\rho}>\varepsilon_{\sigma}$. The
energetic difference between the CDW and SDW indicates the presence of the
spin-charge separation in a LL.

The above discussion implies a hierarchy of energy scales characteristic for
the model considered
\begin{equation*}
2E_{\sigma} <\varepsilon_{\sigma} 
<\varepsilon_0< \varepsilon_{\rho} < E_{\rho}\,. 
\end{equation*}
This suggests to use $E_{\sigma}$ as the natural energy scale. Without
interaction
$2E_{\sigma}=\varepsilon_{\sigma}=\varepsilon_{\rho}=\varepsilon_0$.

\section{Tunneling}

\subsection{Transition rates}
\label{sec:tranrat}

For high tunnel barriers and not too low temperatures, the dominant processes
that contribute to the electron transport are sequential transfers of single
electrons with spin up or down through the two barriers. In this case, $H_{\rm
  t}$ is treated as a perturbation.  Higher order coherent processes can be
safely neglected as long as $k_{\rm B}T\gg \delta E$, with $\delta E$ the
level broadening of virtual states proportional to the tunneling rates.
\cite{fu98} We consider processes that change the state of the dot from an
initial $\left|\rm i\right>$ to a final state $\left|\rm f\right>$,
\begin{equation*}
|{\rm i}\rangle=|n_{\rm i},s_{\rm i},\{l_{q{\rm i}}^{\rho}\},
\{l_{q{\rm i}}^{\sigma}\}\rangle\!,\,\,\,\,\,
|{\rm f}\rangle=|n_{\rm f},s_{\rm f},\{l_{q{\rm f}}^{\rho}\},
\left\{l_{q{\rm f}}^{\sigma}\right\}\rangle\,.
\end{equation*}
Because of the sequential nature of the tunnel processes these states have to
be consistent with the charge and spin selection rules
\begin{equation}
\label{selectionrule}
\Delta n\equiv n_{\rm f}-n_{\rm i}=\pm 1\,,\,\,\,\,\,
\Delta s\equiv s_{\rm f}-s_{\rm i}=\pm 1\,.
\end{equation}
For obtaining explicitly the tunneling rates $\Gamma^{(\lambda)}_{\left|{\rm
      i}\right>\to\left|{\rm f}\right>}$ across the left ($\lambda={\rm L}$)
and right barriers ($\lambda={\rm R}$) we sum over all possible final lead
states and perform a thermal average over the initial states with the chemical
potentials $\mu_{\lambda}$. From (\ref{eq:tunham}) one finds
\begin{equation}
\label{eq:ratekinaret}
\Gamma^{(\lambda)}_{\left|{\rm i}\right>\to\left|{\rm f}\right>}
= t_{\lambda}^{2}\varphi_{\rm d}^{\lambda}\int_{-\infty}^{\infty}{\rm d}\tau\
e_{}^{i \Delta {\mathcal U} \tau} e_{}^{-W_{\rm l}^{}(\tau)}\,.
\end{equation}
The matrix element $\varphi_{\rm d}^{\lambda}$ is  
\begin{equation}
\varphi^{\lambda}_{\rm d} = \left|\left<{\rm i}\right|
\psi_{s,\rm r}^{({\rm d})}
(x^{}_{\lambda},0)\left|{\rm f}\right>\right|^2\, .
\end{equation}
The energy difference
\begin{equation}
\Delta {\mathcal U} = 
{\mathcal U}\left(\left|{\rm  f}\right>\right)-{\mathcal
  U}\left(\left|{\rm i}\right>\right)-
\Delta n\left[\frac{\delta
    C}{C_{\Sigma}^{}}\mp 1\right]\frac{eV}{2}
\end{equation}
is obtained from ${\mathcal U}$ defined in (\ref{eq:confen}). The signs $\pm$
refer to the left ($+$) or right ($-$) barriers, the factor
$\exp{\left[-W_{\rm l}^{}(t)\right]}$ results from the trace over the lead
excitations,\cite{fu98}
\begin{equation}
e^{-W_{\rm l}^{}(\tau)}=\left< 
\psi_{s,\rm r}^{(\lambda)\dagger}(x^{}_{\lambda},\tau)
\psi_{s,\rm r}^{(\lambda)}
(x^{}_{\lambda},0)\right>_{\rm leads}\,.
\end{equation}
The thermal average is performed with respect to the decoupled Hamiltonians
$H_{0}^{(\lambda)}$. This factor turns out to be independent of the spin of
the tunneling electron and the barrier variables. The contribution of the
quantum dot is contained in $\Delta{\mathcal U}$ and in the matrix elements
$\varphi_{\rm d}^{\lambda}$.

The tunneling rates $\Gamma^{(\lambda)}_{\left|{\rm i}\right>\to\left|{\rm
      f}\right>}$ were recently used by Kim {\em et al}.\cite{kinaret} in
order to study the out-of-equilibrium dynamics in the presence of an external
voltage. In this work it has been assumed that the collective modes are
stable and do not relax during the tunneling processes. The probability
distribution was obtained self-consistently using a generalized master
equation.

The opposite situation has been addressed by Braggio {\em et al}.\cite{b01}
who considered the possibility of fast relaxation of the collective modes
induced by extra processes\cite{nazarov2} that are not included in the
diagonal Hamiltonians $H_{0}^{(\rm d)}$, such as spin-orbit interaction and
phonon coupling. In this case, the
dynamical variables consist of the total charge and spin only, while
the collective modes are traced out with a thermal average over the
initial states and a summation over the final
states. In the following, we use this second model. 

As already pointed out\cite{kinaret}, the two approaches do not give
qualitatively different results regarding the current-voltage characteristic
of the system. However, in a very recent work,\cite{kinaret2} the presence of
stable plasmons in a spinless 1D quantum dot has been shown to affect
dramatically its shot noise properties. On the contrary it was also
predicted a crossover towards the results of a model with fully
relaxed bosonic modes\cite{branoise}, in the presence of a
phenomenological relaxation rate for the plasmons. In particular the
two models give the same results if $\gamma_{\rm
p}/\tilde{\Gamma}_0\gtrsim 1$, where $\gamma_{\rm p}$ is a
phenomenological relaxation rate of the plasmonic modes and
${\tilde{\Gamma}_0}^{-1} = {\Gamma_0^{\rm (L)}}^{-1}+{\Gamma_0^{\rm
(R)}}^{-1}$ with $g_{\rho}=g_{\sigma}=1$ - see Eq. \ref{eq:defg0} - is
a characteristic tunneling rate. In a
system with spin, we can expect a similar result if the
phenomenological relaxation rate of the spin density waves
$\gamma_{\rm s}$, satisfies $\gamma_{\rm s}/\tilde{\Gamma}_0 \gtrsim
1$.  For
semiconductor-based 1D quantum dots\cite{a00} one can estimate $\tilde{\Gamma}_0\approx 10^{11}$ ${\rm
s}^{-1}$ by using a tunnel resistance $R_{\rm R} \approx
100\ h/{e}^2$ and $a\approx 0.2\ \mu{\rm m}$ - see section \ref{concl}.
. It is difficult to evaluate
microscopically the relaxation rates $\gamma_{\rm p,s}$. However, one can
estimate,\cite{nazarov2} using a level spacing of about 1 meV, $\gamma_{\rm
  s}\approx 8\cdot 10^{12}$ $s^{-1}$. It can be expected that $\gamma_{\rm
  p}$ is even larger. In the following we assume $\gamma_{\rm
  p,s}/\tilde{\Gamma}_{0} > 1$ and thus complete relaxation of the bosonic
modes.

The reduced rates are then given by
\begin{equation}
\label{ratebb}
\Gamma^{(\lambda)}_{\left|n_{\rm i},s_{\rm i}\right>\to 
\left|n_{\rm f},s_{\rm f}\right>}=
\sum_{\{l_{q{\rm i}}^{\nu}\}}P(\left\{l^{\nu}_{q{\rm i}}\right\})
\sum_{\{l_{q{\rm f}}^{\nu}\}}
\Gamma^{(\lambda)}_{\left|{\rm i}\right>\to\left|{\rm f}\right>}
\end{equation}
with $P(\left\{l^{\nu}_{q{\rm i}}\right\})$ the thermal probability
distribution with respect to $H_0^{(\rm d)}$. Performing the sums one finds
\begin{equation}
\label{eq:ourrate}
\Gamma^{(\lambda)}_{\left|n_{\rm i},s_{\rm i}\right>\to 
\left|n_{\rm f},s_{\rm f}\right>}=
t_{\lambda}^{2}\int_{-\infty}^{\infty}{\rm d}\tau\ e_{}^{i \Delta U \tau} 
e_{}^{-W_{\rm l}^{}(\tau)} 
e_{}^{-W_{\rm d}^{}(\tau)}
\end{equation}
where 
\begin{eqnarray}
\Delta U &=&\frac{E_{\rho}^{}}{2}\left[1+2\left(n_{\rm i}
      ^{}-n_{\rm
      g}^{}\right)\Delta n\right]+
\frac{E_{\sigma}^{}}{2}\left[1+2s_{\rm i}^{}\Delta s\right]
\nonumber\\
&&\nonumber\\
&&\qquad\qquad\qquad\qquad -\Delta n\left[\frac{\delta
    C}{C_{\Sigma}^{}}\mp 1\right]\frac{eV}{2}
\end{eqnarray}
is the energy difference associated to the particular process, the signs $\pm$
refers to the left ($+$) or right ($-$) barriers. The kernel $e^{-W_{\rm
    d}^{}(\tau)}$ represents the thermal average over the initial spin and
charge collective modes in the dot. Using the bosonization method one
obtains\cite{fabrizio} ($\beta^{-1}=k_{\rm B}^{}T$) 
\begin{eqnarray}
W_{\rm l,d}^{}(\tau)&=&\int\ {\rm d}\omega
\frac{J_{\rm l,d}^{}(\omega)}{\omega^{2}}
\Bigg\{\coth\left(\frac{\beta\omega}{2}\right)
\left[1-\cos{(\omega
    \tau)} \right]\nonumber\\&&\nonumber\\
&&\qquad\qquad\qquad\qquad +i \sin{(\omega \tau)}\Bigg \}\,.
\end{eqnarray}
Here, the spectral densities of the leads and the quantum dot are,
\begin{equation}
\label{eq:specdens}
J_{\rm l}(\omega) = 
\frac{\omega}{g} e^{-{\omega}/{\omega_{\rm c}}}
\end{equation}
and
\begin{equation}
J_{\rm d}(\omega) = 
\omega\sum_{\nu\in\{\rho,\sigma\}} \frac{\epsilon_{\nu}^{}}{2
  g_{\nu}} \sum_{m=1}^{\infty}  \delta(\omega - m
\epsilon_{\nu}) e^{-{\omega}/{\omega_{\rm c}}}\,.
\end{equation}
The cutoff $\omega_{\rm c}$ defines the highest energy in the model and
$g^{-1} = (1+g_{0}^{-1})/2$.

The structure of the spectral density of the dot, $J_{\rm d}^{}(\omega)$,
indicates that, though they are infinitely quickly relaxing, the CDW and the
SDW still contribute to the tunneling dynamics. Even if the relaxation
prevents the collective excitations to be initial states for the tunneling, it
is still possible to reach an excited state as a ``final'' state with a given
energy. A typical process is
\begin{equation}
|n_{\rm i},s_{\rm i}\rangle \to 
|n_{\rm f},s_{\rm f},\{l_{q{\rm f}}^{\rho}\},
\{l_{q{\rm f}}^{\sigma}\}\rangle
\rightrightarrows |n_{\rm f},s_{\rm f}\rangle
\end{equation}
where the rightmost process ($\rightrightarrows$ arrow) is associated with a
fast time scale and the intermediate state contains collective excitations.
In this approximation, only the energy of the collective excitation is
detectable. In the following we will use the notation
\begin{equation}
\left|n,s,l_{\rho},l_{\sigma}\right>
\end{equation}
to label an excited state when it is involved as a final state in a tunneling
process.

Let us now investigate in more detail the energy dependence of the rates. For
simplicity, we do not specify the initial and final spin and charge states.
Exploiting the discrete nature of the dot spectral density we rewrite
(\ref{eq:ourrate}) as a function of the energy difference $E=\Delta U$
\begin{equation}
\label{eq:ratesum}
\Gamma^{(\lambda)}_{} (E) = \Gamma^{(\lambda)}_{0}  
\sum_{l_{\rho}^{},l_{\sigma}^{}}  a_{l_\rho} a_{l_\sigma} \gamma
\left(E- l_{\rho}^{}\varepsilon_{\rho}^{} - l_{\sigma}^{}
  \varepsilon_{\sigma}^{} \right)\, ,
\end{equation} 
where
\begin{equation}
\label{eq:defg0}
{\Gamma}^{(\lambda)}_{0}= {\left(\frac{\varepsilon_{\rho}^{}}{\omega_{\rm
        c}^{}}\right)}^{1/2g_{\rho}^{}}{\left(\frac{\varepsilon_{\sigma}^{}}
{\omega_{\rm c}^{}}\right)}^{1/2g_{\sigma}^{}}\frac{2 \omega_{\rm c}^{}
G_{\lambda}^{}}{e^2 \Gamma(1+\alpha)}\, ,
\end{equation}
with 
\begin{equation}
  \label{eq:alpha}
  \alpha=\frac{1}{g}-1
\end{equation}
and $G_{\lambda} = R_{\lambda}^{-1} = \pi e^2 t_{\lambda}^{2} / \omega_{\rm
  c}^{2}$ the intrinsic conductances of the barriers.  The function
$\gamma(x)$ is determined by the leads\cite{b01}
\begin{equation}
\label{eq:gam1}
\gamma(x) = \frac{1}{2 \pi}e^{\beta x/2}
\left|\Gamma\left(\frac{1}{2 g}+i\frac{\beta x}{2\pi}\right)\right|^2 
{\left(\frac{2\pi}{\beta\omega_{{\rm c}}} \right)}^{\alpha}\, .
\end{equation}
At $T=0$ we have
\begin{equation}
\label{eq:gam2}
\gamma^{0}(x) = \left(\frac{ {x}}{\omega_{{\rm c}}}\right)^{\alpha}
\theta(x)\,.
\end{equation}
The weights $a_{l_\nu}$ are due to the dot contribution. At $T=0$, they are
\begin{equation}
\label{eq:azero}
a_{l_\nu}^{0} = \frac{\Gamma\left(1/2
      g_{\nu}+l_{\nu}^{}\right)}{\Gamma\left(1/2
      g_{\nu}\right) l_{\nu}^{}!} \theta(l_{\nu}^{})\,.
\end{equation}
For finite temperature, the weights have to be numerically determined. In
(\ref{eq:gam1})-(\ref{eq:azero}), $\Gamma(z)$ is the Euler-gamma function.
Figure~{\ref{fig:figrates} shows the effect of the charge-spin separation in
  the structure of $\Gamma^{(\lambda)}_{}(E)/\Gamma_{0}^{(\lambda)}$ at $T=0$.
\begin{figure}[htbp]
  \begin{center}
    \includegraphics[width=6.5cm,keepaspectratio]{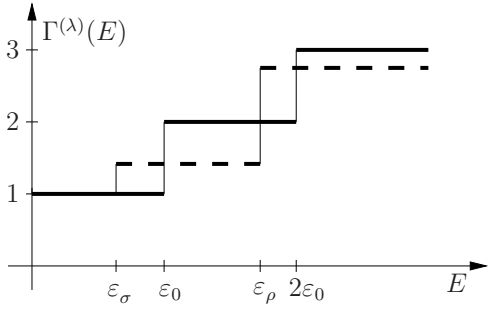}
    \caption{Schematic plot of the transition rate $\Gamma^{(\lambda)}_{}(E)$, 
      in units of $\Gamma_{0}^{(\lambda)}$, as a function of the tunneling
      energy $E$, for $g_{0}= 1$ and $T=0$. Full line: no interactions in the
      dot, $g_{\rho}=g_{\sigma}=1$; dashed line: with interactions,
      $g_{\sigma} > 1$ and $g_{\rho} < 1$. Excitation energies are indicated
      at the energy-axis.}
    \label{fig:figrates}
  \end{center}
\end{figure}
In the absence of spin-charge separation ($g_{\rho}=g_{\sigma}=1$) all of the
energies are degenerate,
$\varepsilon_{\rho}=\varepsilon_{\sigma}=\varepsilon_0$. It is then easy to
write down explicitly the contributions to the sum ({\ref{eq:ratesum})
\begin{equation*}
\begin{split}
&{a}_{0}^{2}\theta(x)+2{a}_{0}^{}{a}_{1}^{}\theta(x-\varepsilon_0)
+\left({a}_{1}^{2}+2{a}_{0}^{}{a}_{2}^{}\right)\theta(x-2\varepsilon_0)\\
&+2\left({a}_{0}^{}{a}_{3}^{}+{a}_{1}^{}{a}_{2}^{}\right)
\theta(x-3\varepsilon_0)+\ldots
\end{split}
\end{equation*}
with ${a}_{0}=1$, ${a}_{1}=1/2$, ${a}_{2}=3/8$, ${a}_{3}=5/16$, $\ldots$. Each
of the coefficients of the theta-functions in the latter expression sums up to
one (Fig.~\ref{fig:figrates}), so one always has unit steps centered at
integer values of $\varepsilon_0$. If $g_{\sigma} > 1$ and $g_{\rho} < 1$,
charge and spin modes are energetically split,
$\varepsilon_{\sigma}<\varepsilon_0< \varepsilon_{\rho}$. Then, the first
three terms of (\ref{eq:ratesum}) are
\begin{equation*}
\theta(x)+\frac{1}{2g_{\sigma}}\theta(x-\varepsilon_{\sigma})
+\frac{1}{2g_{\rho}}\theta(x-\varepsilon_{\rho})\,.
\end{equation*}
This implies non-integer quantized steps at different energies with heights
depending on the spin and charge interactions (Fig.~\ref{fig:figrates}, dashed
line). For simplicity, in Fig.~{\ref{fig:figrates} $T=0$ and $g_0=1$
have been assumed. Finite temperatures and $g_0<1$ do not
  drastically change these results but smoothen the jumps at the positions
  given by the dot parameters.

\subsection{The master equation}
\label{mastereq}

Having assumed fast relaxation of the collective excitations in the quantum
dot necessarily implies that the stationary states of the dot are fully
characterized by the variables $n$ and $s$.  One can then define an occupation
probability $P(n,s)$ for a state $|n,s\rangle$ that satisfies the master
equation
\begin{eqnarray}
\label{eq:meq1}
\partial_{t} P_{n,s}(t) =& \sum_{\substack{n'=n\pm1\\s'=s\pm1}}\left[
  P_{n',s'}(t) \Gamma_{\left|n',s'\right>\to\left|n,s\right>}\right.
\nonumber\\
& -\left. P_{n,s}(t)
  \Gamma_{\left|n,s\right>\to\left|n',s'\right>}\right]\, , 
\end{eqnarray}
where $\Gamma \equiv \sum_{\lambda} \Gamma^{(\lambda)}_{}$. In the stationary
limit, the l.h.s. of (\ref{eq:meq1}) is zero such that one has to solve a
homogeneous system of linear equations. The solution must be normalized,
$\sum_{n,s} P_{n,s} = 1$. The stationary current is
\begin{eqnarray}
\label{eq:current}
I = & e \sum_{n,s}\sum_{q=\pm1} P_{n,s}
\left[\Gamma_{\left|n,s\right>\to\left|n+1,s+q\right>}^{({\rm
      R})}\right.\nonumber\\
&\left.-\Gamma_{\left|n,s\right>\to\left|n-1,s+q\right>}^{({\rm R})}\right]\, .
\end{eqnarray}
If both $eV$, $k_{{\rm B}}T < E_{\rho}^{}$ at most two charge states enter the
dynamics, and the current is given by tunneling events corresponding to
transitions
$\left|n,s\right>\to\left|n+1,s'\right>\to\left|n,s''\right>$. Below,
we will consider this regime.

In the above discussion, it is assumed that states with spin $|s| > 1$ are
stable. Spin-flip processes, however, can induce a relaxation of the total
spin. These can be due to different mechanisms such as magnetic scattering and
spin-orbit interaction. Another possible source of spin-flip processes can be
cotunneling. The quantitative microscopical evaluation of the corresponding
relaxation rates is not easy and depends on the dimensionality of the system
and on the electronic correlations. In GaAs-based quantum dots, it seems that
spin-flip processes have much smaller rates\cite{nazarov2}, four orders of
magnitude or even smaller than the non-spin-flip processes involved in the
relaxation of the bosonic modes. In 1D systems, the non-Fermi liquid nature of
the interaction could lead to a non-trivial energy dependence of these
relaxation rates.

It is primarily not the aim of the present paper to study these processes. In
order to get first insight into the stability of the negative differential
conductance features it is sufficient to introduce a {\em phenomenological}
spin-flip relaxation rate $\Gamma_{s\to s'}^{w}$, with $s'=s\pm 2$, and modify
accordingly Eq. (\ref{eq:meq1}). The results of this modification will be
discussed in section \ref{sec:phasediag}.

\section{Linear transport}

In this section we specialize (\ref{eq:current}) to the linear regime ($V \to
0$). We assume symmetric barriers, $t_{\rm L} \equiv t_{\rm R}$, and $C_{\rm
  L}\equiv C_{\rm R}$, hence we drop for convenience the barrier index in the
expression for the rates. The electrochemical potential of the quantum dot is
defined as
\begin{equation}
\mu_{{\rm d}}^{}(n,s_{n}^{}) = {\mathcal U}(n+1,s_{n+1}^{},0,0)-{\mathcal
  U}(n,s_{n},0,0)
\end{equation}
where $s_{n}$, $s_{n+1}$ are the spins of the ground state for $n$, $n+1$
electrons, respectively. From (\ref{eq:confen}) one finds
\begin{equation*}
\mu_{{\rm d}}(n,s_{n}^{}) =  E_{\rho}^{}
\left(n+\frac{1}{2}-n_{g}^{}\right)+(-1)_{}^{n}\frac{E_{\sigma}^{}}{2}\, .
\end{equation*}
In the following, we evaluate the linear differential conductance  
$G = \left.\partial I/\partial V\right|^{}_{V=0}$.

\subsection{Low temperature}
\label{sec:ltzt}

At low temperatures, $k_{\rm B}^{}T \ll E_{\sigma}^{}$, the master equation is
solved in the subspace defined by the ground states for $n$ and $n+1$
electrons. The conductance is
\begin{equation}
\label{eq:theconductance}
G(\xi) = \frac{\beta e^2}{\sqrt{8}}
\frac{\Gamma(\xi)
  e^{-\beta\xi/2}}{\cosh\{ [\beta\xi+(-1)^{n}\log 2]/2\}}
\end{equation}
where $\Gamma(\xi)$ is the rate in (\ref{eq:ratesum}) and $\xi = E_{\rho}
\left[n_{g}^{}-n_{g}^{{\rm res}}(n,0)\right]$ corresponds to the deviation
from the resonance peak position $n_{g}^{{\rm res}}(n,0)$ at $T=0$. The latter
is determined by the condition $\mu_{\rm d}=0$
\begin{equation}
n_{g}^{{\rm res}}(n,0) = n + \frac{1}{2} + (-1)^{n}
\frac{E_{\sigma}^{}}{2 E_{\rho}^{}}\, .
\end{equation}
The position of the linear conductance peaks is affected by the spin that
leads to an even-odd effect in the distances $\delta(n
\leftrightarrow n+1)$ between conductance peaks
\begin{equation*}
\delta(n \leftrightarrow n+1) = 1 + (-1)^{n+1}
\frac{E_{\sigma}}{E_{\rho}}\, .
\end{equation*}
This even-odd effect was recently observed in carbon nanotube
experiments.\cite{cetal98}

For $T\neq 0$ the peaks in the conductance no longer occur at the zero
temperature positions. Instead, they shift linearly with temperature with
slopes that depend on the interaction in the leads (Fig.~\ref{fig:slope})
\begin{equation}
\label{eq:peakpos}
n_{g}^{{\rm res}}(n,T) = n_{g}^{{\rm res}}(n,0)+{(-1)}^{n+1}
\phi(g_{0}^{}) \frac{k_{{\rm B}}^{}T}{E_{\rho}^{}}\, .  
\end{equation}
Here, $\phi$ is obtained from the implicit relation
\begin{equation}
{\rm
  Im}\left[\psi\left(\frac{1}{2 g}\!+\!i
\frac{\phi(g_{0})}{2\pi}\right)\right]\!
+\!\frac{\pi}{2}{\rm tanh}\left[\frac{\log 2\!+\!
\phi(g_{0})}{2}\right]\!=0
\end{equation}
with $\psi(z)$ the digamma-function.  Without interaction in the leads
($g_{0}^{}=1$) one recovers the well know result $\phi(1)=\log
2/2$.\cite{baranger}
\begin{figure}[htbp]
  \begin{center}
    \includegraphics[width=7cm,keepaspectratio]{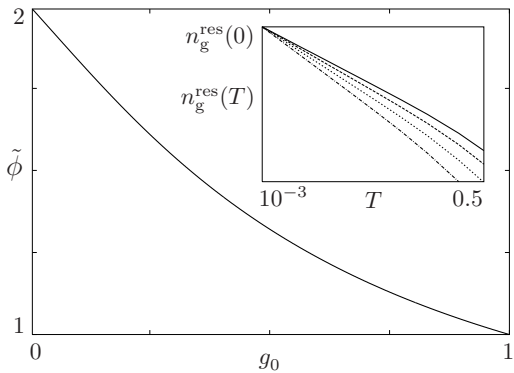}
    \caption{The prefactor
      $\tilde{\phi}=2\phi\left(g_{0}^{}\right)/\log 2$ in (\ref{eq:peakpos})
      as a function of the interaction in the leads.  Inset: low-temperature
      shift of a conductance peak for $g_{0}^{} = 1.0$ (solid), $g_{0}^{} =
      0.8$ (dashed), $g_{0}^{} = 0.6$ (dotted), $g_{0}^{} = 0.4$ (dash-dotted)
      (temperature units $E_{\sigma}/k_{\rm B}$).}
    \label{fig:slope}
  \end{center}
\end{figure}

For $ k_{\rm B}^{} T \ll \varepsilon_{\nu}^{}$, since one cannot excite charge
and spin density waves in the dot, the temperature dependence of the
conductance maximum is determined only by the interactions in the leads
\begin{equation}
\label{eq:plawlowT}
G_{}^{\rm max}(T)\propto T^{\,\alpha-1}_{}
\end{equation}
with $\alpha$ given in (\ref{eq:alpha}).

\subsection{High temperature}

For $E_{\sigma}, \varepsilon_{\sigma} < k_{{\rm B}}T$ the peak position
deviates from the linear behavior and approaches, for $k_{\rm B}T \approx
E_{\sigma}$, the spinless value $n + 1/2$ (Fig.~\ref{fig:peakpos}).
\begin{figure}[htbp]
  \begin{center}
    \includegraphics[width=7cm,keepaspectratio]{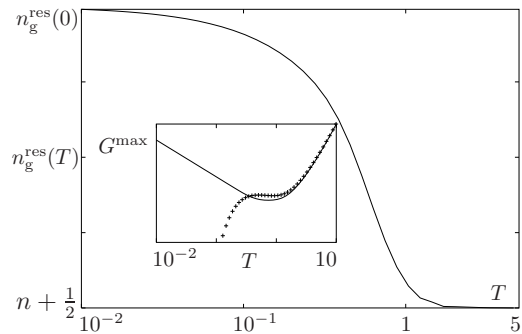}
    \caption{ Position of the conductance
      peak for even $n$, as a function of the temperature $T$ (units
      $E_{\sigma}/k_{\rm B}$) for $g_{0}^{}= 0.8$, $g_{\rho} = 0.3$ and
      $g_{\sigma} = 1.0$. At high temperatures the peak position approaches
      the spinless value $n+1/2$. Inset: double logarithmic plot of the
      conductance maximum ${G}^{\rm max}(T)$ in arbitrary units. Crosses:
      analytic result from (\ref{eq:asympt}).}
    \label{fig:peakpos}
  \end{center}
\end{figure}
For $k_{\rm B}T \gg E_{\sigma}^{}$, one obtains an analytic
expression for the conductance. We factorize the occupation probability,
$P_{n,s} = p(n) \Psi(s)$ with $\Psi(s)$ the thermal occupation probability
\begin{eqnarray}
\label{eq:ansatz}
\Psi(s)=\left\{\begin{array}{ll}
Z_{\rm e}^{-1} e^{-{\beta E_{\sigma} s^2}/2}&\qquad\text{$s$ even}\\
Z_{\rm o}^{-1} e^{-{\beta E_{\sigma} (s^2-1)}/2}&\qquad\text{$s$ odd}
\end{array}\right.
\end{eqnarray}
The prefactors $Z_{\rm e}=Z_{\rm o} = \sqrt{\pi/2\beta E_{\sigma}}$ are
determined by the normalization conditions
$\sum_{s}^{}\Psi(2s)=\sum_{s}^{}\Psi(2s+1) =1$.  Using the factorized form for
$P_{n,s}$ we sum over the spin states in the master equation obtaining
$p(n) \tilde{\Gamma}_{n\to n+1} = p(n+1) \tilde{\Gamma}_{n+1\to n}$
($n$ even) with
\begin{equation}
\tilde{\Gamma}_{n\to n+1} =\!\!\sum\limits_{p=\pm 1}\int {\rm d}s\ e^{-2\beta
  E_{\sigma}^{} s^2}
\Gamma_{\left|n,2s\right>\to\left|n+1,2s+p\right>}\,,\nonumber
\end{equation}
and
\begin{equation}
\tilde{\Gamma}_{n+1\to n} =\!\!\sum_{p=\pm 1}\int{\rm d}s\ e^{-2\beta
  E_{\sigma}^{} (s^2 + p s)}
\Gamma_{\left|n+1,2s+p\right>\to\left|n,2s\right>}\,.\nonumber
\end{equation}
The final result for the differential conductance is
\begin{equation}
\label{eq:asympt}
{G}(\zeta) = \frac{\beta e^2}{2} 
\frac{ \Gamma\left( \zeta + E_{\sigma}/2\right)
\Gamma\left( -\zeta - E_{\sigma}/2\right) }
{ \Gamma\left( \zeta + E_{\sigma}/2\right)
+\Gamma\left( -\zeta - E_{\sigma}/2\right)}\, ,
\end{equation}
with $\zeta= E_{\rho}^{}\left(n -n_{g}^{}+1/2\right)$.  When $k_{\rm B}^{}T
\gg E_{\sigma}^{}$ the conductance peak as extracted from (\ref{eq:asympt}) is
centered around $\zeta=0$ which implies $n_{g}^{{\rm res}}(n,T) = n + 1/2$,
the spinless position. The conductance peak maximum however follows different
power laws depending on the temperature range considered
\begin{eqnarray}
\label{eq:ansatz1}
G^{\rm max}(T)\propto\left\{\begin{array}{ll}
T^{\,\alpha-1+\left(2g_{\sigma}\right)^{-1}_{}}
&\quad\text{$\varepsilon_{\sigma}^{} \ll 
k_{\rm B}^{}T < \varepsilon_{\rho}^{}$}\\
T^{\,\alpha-1+\left(2g_{\sigma}\right)^{-1}_{}+
\left(2g_{\rho}\right)^{-1}_{}}&\quad
\text{$\varepsilon_{\rho}^{} \ll k_{\rm B}^{}T<E_{\rho}$}
\end{array}\right.\nonumber
\end{eqnarray}
Thus, for a well developed energetic separation between charge and spin it
should be in principle possible to detect three different power laws in the
linear conductance peak maximum value, over the whole temperature range.

\section{Nonlinear transport}

In the following, we discuss the transport in the nonlinear regime.
\begin{figure}[htbp]
\setlength{\unitlength}{1cm} 
\begin{center}
\includegraphics[width=7cm,keepaspectratio,clip]{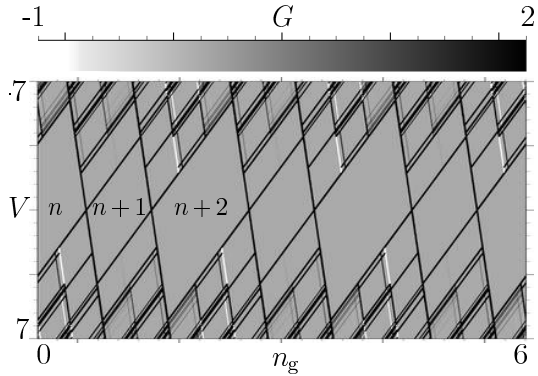}
    \caption{The differential conductance $G$ 
      (arbitrary units) as a function of bias voltage $V$ (units
      $E_{\sigma}/e$) and number of gate charges $n_{\rm g}$. System
      parameters are $A= G_{\rm L}^{}/G_{\rm R}^{}=50$, $k_{\rm
        B}T=10^{-2}E_{\sigma}$, $C_{\rm L}=5C_{\rm R}$, $C_{\rm g}/C_{\rm R} =
      0.01$ $g_{\rho} =0.63$, $g_{\sigma} = 1.15$, $g_{0}=1$,
      $E_{\rho}=5E_{\sigma}$. Top: gray scale.  }
    \label{fig:manyn}
  \end{center}
\end{figure}
We assume $n$ to be even, then we label the states $\left|n,s\right>$ using
the spin variable $s$ alone such
$\Gamma_{\left|n,s\right>\to\left|n',s'\right>}^{(\lambda)}\,\to \Gamma_{s\to
  s'}^{(\lambda)}$. The master equation for the stationary probabilities
$P_{n,s}$ is solved numerically. From this the current-voltage characteristic
$I(V,n_{\rm g}^{})$ and the differential conductance $G(V,n_{\rm g}^{}) =
\partial I/\partial V$ are calculated. Figure~\ref{fig:manyn} gives an
overview of the behavior of the differential conductance as a function of bias
voltage $V$ and gate voltage induced number of charges $n_{\rm g}$. Most
pronounced are the parity effect for $V\to 0$ and the rich structure due to
the excited states of the quantum dot. This can be understood in detail by
considering all the possible transitions and taking into account the
correlation induced features of the tunneling rates. Most important are the
negative differential conductance peaks indicated by the white lines.  They
are closely related to the non-Fermi liquid properties of the model.  This is
discussed below in more detail.

\subsection{Symmetric barriers}
\label{herediscuss}

Figure \ref{fig:Gsymm} shows the differential conductance $G(V,n_{\rm g})$ for
symmetric barriers as a function of bias and gate voltages.  The conductance
is always positive.  Each of the lines corresponds to a peak in the
differential conductance and is related to a transition between states in the
dot for $n$ and $n+1$.
\begin{figure}[htbp]
  \begin{center}
    \includegraphics[width=7cm,keepaspectratio]{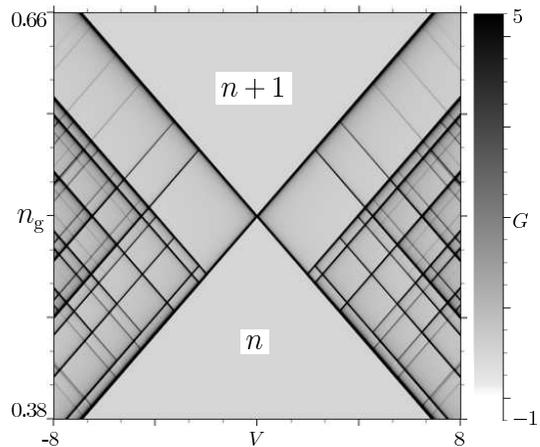}
    \caption{The differential conductance $G$ 
      (arbitrary units) as a function of the bias voltage (units
      $E_{\sigma}^{}/e$) and $n_{\rm g}^{}$, for symmetric barriers ($A=1$,
      $\eta=1/2$) for $k_{\rm B}T=10^{-2}E_{\sigma}$, $E_{\rho}=25E_{\sigma}$, $g_{0}^{} =
      0.9$, $g_{\rho} = 0.8$, $g_{\sigma} = 1.15$ corresponding to
      $\varepsilon_{\sigma}^{}/\varepsilon_{0}^{} = 0.99$,
      $\varepsilon_{\rho}^{}/\varepsilon_{0}^{} = 1.42$ and
      $2E_{\sigma}^{}/\varepsilon_{0}^{} = 0.86$.
      Right: gray scale.}
    \label{fig:Gsymm}
  \end{center}
\end{figure}

The condition for opening the transition
$\left|n,s\right>\leftrightarrow\left|n+1,s\pm 1\right>$, without involving
CDW and SDW, is
\begin{equation}
\label{eq:condspec}
-\eta eV\leq-E_{\rho}^{}\tilde{n}_{\rm g}^{}\pm
E_{\sigma}^{}s\leq(1-\eta) eV\, .
\end{equation}
with $\tilde{n}_{\rm g}^{} = n_{\rm g}^{}-n_{\rm g}^{\rm res}(n,0)$ and
\begin{equation*}
\eta=\frac{C_{\rm R}+C_{\rm g}/2}{C_{\rm L}+C_{\rm R}+C_{\rm g}}\, .
\end{equation*}
For symmetric barriers $\eta = 1/2$ and $A = G_{\rm L}^{}/G_{\rm R}^{} = 1$.
\begin{figure}[htbp]
  \begin{center}
    \includegraphics[width=6.5cm,keepaspectratio]{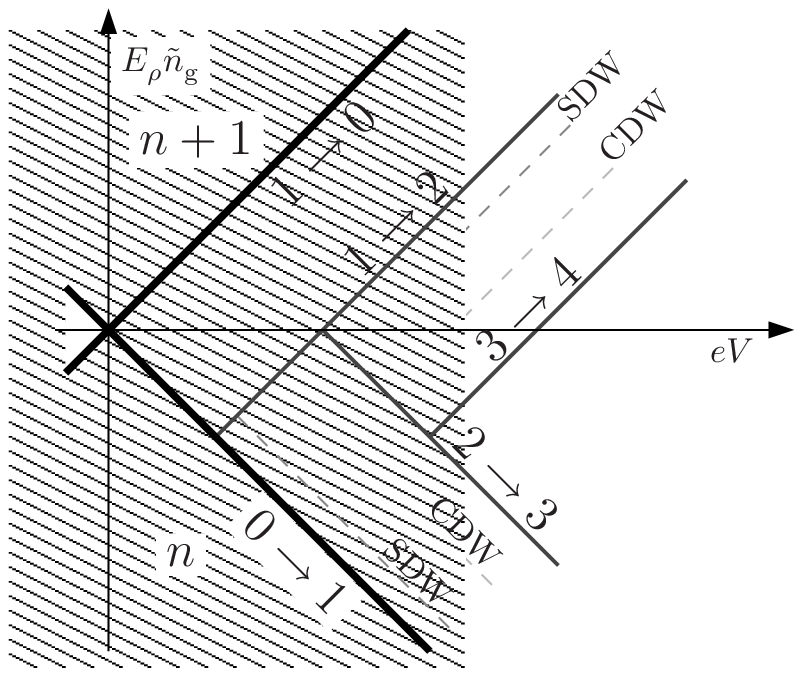}
    \caption{
      Scheme of transition lines in the differential conductance for symmetric
      barriers ($\eta = 1/2$, $A=1$). Shaded regions denote Coulomb blockade.
      Thick black lines: ground-state to ground-state transitions; dark-gray
      sequence of lines: transitions to higher-spin states (spin values of the
      states involved in the transition are indicated); dashed lines: SDW and
      CDW excited states. For simplicity we have dropped indices $n$, $n+1$
      denoting the transitions by spin values $s$ only.}
    \label{fig:struct}
  \end{center}
\end{figure}
The scheme of the transitions consistent with (\ref{eq:condspec}) is shown in
Fig.~\ref{fig:struct}.  States with spin $s$ and $-s$ are degenerate (cf.
(\ref{eq:confen})).  First, we consider transitions between the ground states,
$\left|n,0\rangle\leftrightarrow|n+1,1\right>$ associated with the lines
$E_{\rho} \tilde{n}_{\rm g} = (\eta-1) eV$ and $E_{\rho} \tilde{n}_{\rm g} =
\eta eV$, respectively. These lines divide the plane $\left(V > 0,n_{\rm
    g}^{}\right)$ in three regions. The shaded regions labeled $n$ and $n+1$
denote the Coulomb blockade regimes with zero conductance. In the other
regions, both of the ground states have finite occupation probabilities and
the quantum dot is conducting.

Since the state $\left|n+1,1\right>$ is now occupied, the transition
$\left|n+1,1\right>\to \left|n,2\right>$ becomes available at sufficiently
high voltages. The corresponding activation threshold is given by
$E_{\rho}^{}\tilde{n}_{\rm g}^{}=\eta eV - 2E_{\sigma}^{}$ (cf.
(\ref{eq:condspec}) with $s = 2$).  The transition channel
$\left|n,2\right>\to\left|n+1,1\right>$ is always open in the transport
domain. Inside the domain where $\left|n,2\right>$ is occupied we can achieve
the transition for $\left|n,2\right>\to\left|n+1,3\right>$ by increasing $V$.
By iterating the procedure with increasingly higher values of the spin we get
the fish-bone pattern shown in Fig.~\ref{fig:struct}.

So far we only have considered the excited states with higher spin of the dot.
In order to complete the picture, we must include all the transitions
involving collective charge and spin excitations. This enhances considerably
the complexity of the spectrum, since at high enough bias voltage each
transition of the type $\left|n,s\right>\to\left|n',s'\right>$ can also occur
via the channels $\left|n,s\right>\to\left|n',s',l_{\rho},l_{\sigma}\right>
\rightrightarrows\left|n',s'\right>$. For example, consider the CDW- and
SDW-channels for the transition $\left|n,0\right>\to\left|n+1,1\right>$. These
correspond to the equations $E_{\rho}\tilde{n}_{\rm g}=\left(\eta-1\right)
eV+l_{\rho}\varepsilon_{\rho}+l_{\sigma}\varepsilon_{\sigma}$. The lines are
parallel to the line $\left|n,0\right>\to\left|n+1,1\right>$.  Analogously,
including CDW- and SDW-channels in the transition
$\left|n+1,1\right>\to\left|n,0\right>$ gives rise to
$E_{\rho}^{}\tilde{n}_{\rm g}^{}=\eta
eV+l_{\rho}^{}\varepsilon_{\rho}^{}+l_{\sigma} \varepsilon_{\sigma}^{}$ and to
transition lines parallel to the line $\left|n+1,1\right>\to\left|n,0\right>$
(Fig.~\ref{fig:struct}).

\subsection{Non symmetric barriers}
\label{pdcndc}

Figure~\ref{fig:ndc1} shows the 
differential conductance for asymmetry $A=50$.
\begin{figure}[htbp]
  \begin{center}
    \includegraphics[width=7cm,keepaspectratio]{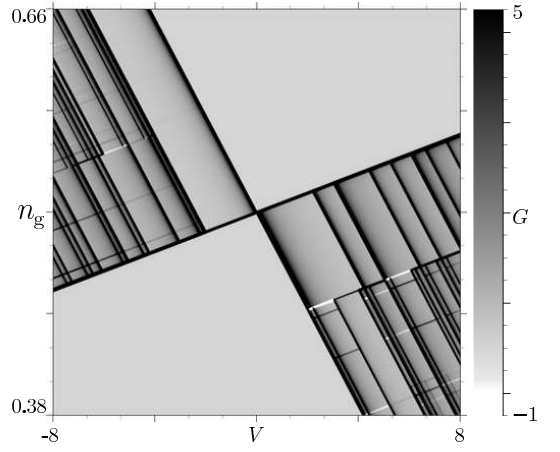}
    \caption{The differential conductance $G$ 
      (arbitrary units) as a function of $V$ (units $E_{\sigma}^{}/e$)
      and $n_{\rm g}$ for asymmetric barriers with $A = 50$, $C_{\rm
        L}^{}=5C_{\rm R}$, $C_{\rm g}/C_{\rm R} = 0.01$, $k_{\rm B}T
      =10^{-2}E_{\sigma}$, $E_{\rho}=25E_{\sigma}$, $g_{0} = 0.9$, $g_{\rho} = 0.8$,
      $g_{\sigma}^{} = 1.1$ corresponding to
      $\varepsilon_{\sigma}^{}/\varepsilon_{0}^{} = 0.995$,
      $\varepsilon_{\rho}^{}/\varepsilon_{0}^{} = 1.37$,
      $2E_{\sigma}^{}/\varepsilon_{0}^{} = 0.9$. Several transitions associated with NDC
      (white lines) parallel to transition lines
      $\left|n+1,2s-1\right>\to\left|n,2s,\rho,\sigma\right>$ are
      observed.  Right: gray scale. }
    \label{fig:ndc1}
  \end{center}
\end{figure}
\begin{figure}[htbp]
  \begin{center}
    \includegraphics[width=7cm,keepaspectratio]{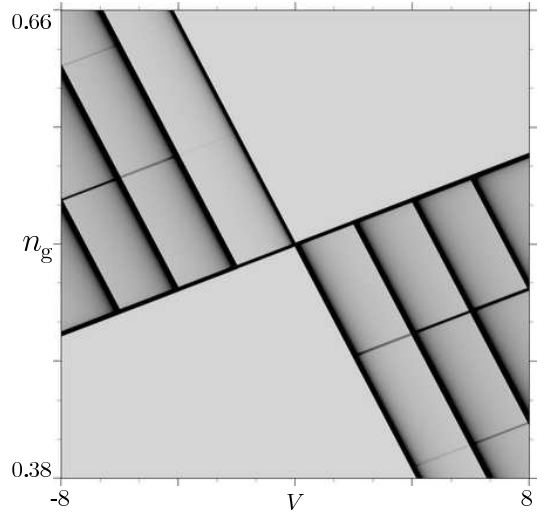}
    \caption{The differential conductance $G$ 
      (arbitrary units) as a function of $V$ (units $E_{\sigma}^{}/e$)
      and $n_{\rm g}$ for $A = 50$, 
      $C_{\rm L}^{}/C_{\rm R} = 5$, $C_{\rm g}/C_{\rm R} = 0.01$,
      $k_{\rm B}T =10^{-2}E_{\sigma}$, $E_{\rho}=25E_{\sigma}$, $g_{0} = 0.9$, $g_{\rho} = g_{\sigma} = 1.0$,  $2E_{\sigma}=
      \varepsilon_{\sigma}=\varepsilon_{\rho}=\varepsilon_{0}$. 
Despite the strong asymmetry there is no NDC. Lines parallel
      to the transitions $\left|n+1,1\right>\to\left|n,0\right>$ tend to have
      very small intensity. Gray scale as in Fig.~\ref{fig:ndc1}.}
    \label{fig:pdc1}
  \end{center}
\end{figure}

The main difference between the results in Fig.~\ref{fig:ndc1} and those
obtained for symmetric barriers (cf. Fig.~\ref{fig:Gsymm}) are the negative conductance features
associated with the white transition lines. They are due to the presence of
interaction, $\varepsilon_{\rho}\not=\varepsilon_{\sigma}$, as can be seen
from Fig.~\ref{fig:pdc1} where the differential conductance is shown for
$g_{\rho} = g_{\sigma} = 1$.

For $V > 0$ electrons flow from right to left. Then, for $A>1$ the electrons
traverse a higher barrier tunneling into the dot, and a lower barrier
tunneling out of the dot. Thus, states $\left|n,s\right>$ will have a higher
occupation probability as compared to states $\left|n+1,s+1\right>$.  This
``trapping'' phenomenon at sufficiently large asymmetries can create a
bottleneck for the electron transport and enhances the probability to have a
NDC.  The occupation probabilities for the lower spin states are shown in
Fig.~\ref{fig:prob} for increasing bias voltage. If the occupation of a
higher-spin state e.g. $(n,2)$ is favored on the expense of the ground state,
$(n,0)$, it accumulates occupation probabality and this eventually leads to a
reduction of the current.

However, it is important to note that trapping alone is not sufficient to
induce NDC. If the system does not exhibit spin-charge separation (Fig.
\ref{fig:pdc1}) no NDC occurs, independent of the strength of the asymmetry.
\begin{figure}[htbp]
\setlength{\unitlength}{1cm}
\begin{center}
  \includegraphics[width=7cm,keepaspectratio]{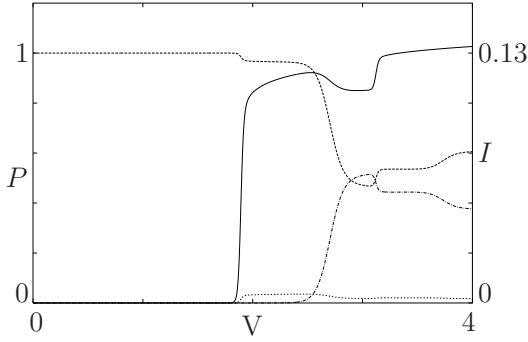}
    \caption{Occupation probabilities $P_{n,0}$ (dashed), $P_{n,2}+P_{n,-2}$ 
      (dashed-dotted), $P_{n+1,1}+P_{n+1,-1}$ (dotted), and current $I$ (full line, units
      $e\Gamma_0^{(\rm R)}$) as a function of the bias voltage $V$ (units
      $E_{\sigma}/e$) for $n_{\rm g}=0.457$; other parameters as in
      Fig.~\ref{fig:ndc1}.}
    \label{fig:prob}
  \end{center}
\end{figure}
To understand this point better, we use a simple model that can be treated
analytically.  We assume $g_{\rho}=g_{\sigma}=g_0=1$ and $k_{\rm B}T
\ll \varepsilon_{\sigma}$.
\begin{figure}[htbp]
  \begin{center}
    \includegraphics[width=7cm,keepaspectratio]{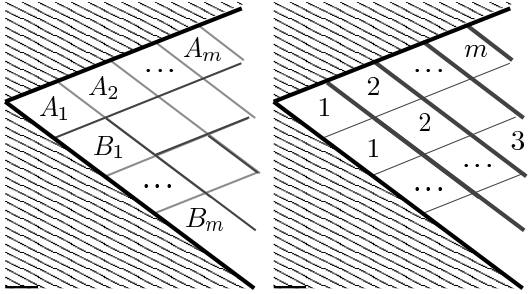}
    \caption{Transport without spin-charge separation. 
      Left: transition lines for asymmetric barriers ($A > 1$)
      and $g_{\rho} = g_{\sigma} = 1$. Right: values for the current ${I}$ in
      units ${I}_{0} = e\Gamma_{0}^{(\rm R)}$ for $A \to \infty$ and $T=0$, keeping
      $G_{\rm R}$ finite.}
    \label{fig:proof}
  \end{center}
\end{figure}

We refer to Fig. \ref{fig:proof} (left) where several regions of given
higher-spin states are shown: $\{A_{m}\}$ with $s_{\max}=1$, $\{B_{m}\}$ with
$s_{\max}=2$. In these regions the transition rates occur in a well defined
pattern. For instance, for $\{A_{m}\}$ we have $\Gamma_{0\to 1}^{(\rm
  R)}=m\Gamma_{0}^{(\rm R)}$ and $\Gamma_{1\to 0}^{(\rm L)}=\Gamma_{0}^{(\rm
  L)}$, while for $\{B_{m}\}$ we have $\Gamma_{0\to 1}^{\rm
  (R)}=\Gamma_{0}^{(\rm R)}$, $\Gamma_{1\to 0}^{(\rm L)}=(m+1)\Gamma_{0}^{(\rm
  L)}$, $\Gamma_{1\to2}^{(\rm L)}=m\Gamma_{0}^{(\rm L)}$ and
$\Gamma_{2\to1}^{(\rm R)}=2\Gamma_{0}^{(\rm R)}$. This allows to solve
the master equation and evaluate the
currents (cf. Fig.~\ref{fig:proof} right) for $A\to\infty$. This limit is
performed by assuming $G_{\rm R}$ finite and corresponds to the most favorable
situation for generating NDC. One finds that the conductance lines parallel to
$\left|n+1,1\right>\to\left|n,0\right>$ vanish. The lines parallel to
$\left|n,0\right>\to\left|n+1,1\right>$ correspond to PDC. Thus, for
$A\to\infty$, the absence of charge-spin separation leads to a landscape of
PDC peaks. The numerical results for $A=50$ (Fig.~\ref{fig:pdc1}) are
consistent with this picture. We have confirmed numerically that the above
results remain valid also for interacting leads, $g_{0} < 1$, and for $T\neq
0$.

\subsection{The five-states model}

Among the transitions in Fig.~\ref{fig:ndc1} which show NDC, we will
concentrate on those in the gray region for the transition
$|n,2\rangle\to|n+1,1\rangle$ (cf. Fig.~\ref{fig:zones}). This is divided into
three regions I, II, III depending on the presence of SDW and CDW states,
$l_{\sigma}=l_{\rho}=0$; $l_{\sigma}=1$, $l_{\rho}=0$; $l_{\sigma}=0$,
$l_{\rho}=1$, respectively. Here, only the five states $\left|n,0\right>$,
$\left|n+1,\pm 1\right>$, $\left|n,\pm 2\right>$ contribute to transport.
\begin{figure}[htbp]
  \begin{center}
    \includegraphics[width=5cm,keepaspectratio]{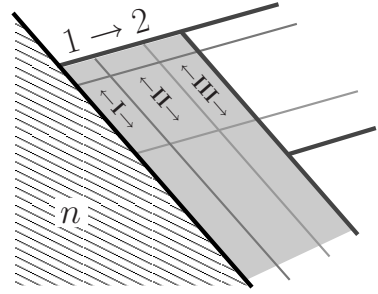}
    \caption{Regions analyzed analytically with respect to the NDC: 
I, II, III correspond to $l_{\sigma}=l_{\rho}=0$; $l_{\sigma}=1$,
      $l_{\rho}=0$; $l_{\sigma}=0$, $l_{\rho}=1$, respectively, in the
      transition $|n,2\rangle\to|n+1,1\rangle$.}
    \label{fig:zones}
  \end{center}
\end{figure}
For these, the master equation is solved analytically. The current is
\begin{equation}
\label{eq:5stcur}
I=e \frac{2\Gamma_{0\to1}^{(\rm R)}\left[\Gamma_{1\to0}^{(\rm
      L)}+\Gamma_{1\to2}^{(\rm L)}\right]\Gamma_{2\to1}^{({\rm R})}}{\Gamma_{1\to0}^{(\rm
   L)}\Gamma_{2\to1}^{(\rm R)}+2\Gamma_{0\to1}^{(\rm
   R)}\left[\Gamma_{2\to1}^{(\rm
     R)}+\Gamma_{1\to2}^{(\rm L)}\right]}\,, 
\end{equation} 
with the transition energies $\Delta U_{0\to1} = eV-2E_{\sigma}$, $\Delta
U_{2\to1}= eV$, $\Delta U_{1\to0}=2E_{\sigma}$ and $\Delta U_{1\to2}=0$ inside
the corresponding rates.  For $k_{\rm B}^{}T \ll E_{\sigma}$ and $g_{0}$ not
too small, one can assume that along the lines
$\left|n+1,1\right>\to\left|n,2\right>$,
$\left|n+1,1\right>\to\left|n,0,0,1\right>$,
$\left|n+1,1\right>\to\left|n,0,1,0\right>$, and not too close to the crossing
points, the only rates that contribute to the derivative with respect to $V$
are $\Gamma_{1\to2}^{(\rm L)}$ and $\Gamma_{1\to0}^{(\rm L)}$. With this the
differential conductance is
\begin{equation}
\label{eq:conductance}
G=\frac{e\phi_{0}^{}}{{\mathcal D}_0^2}\sum_{p=\pm
    1}^{}\Lambda_p\partial_V\Gamma_{1\to1+p}^{(\rm L)}(V)\,,
\end{equation}
where $\phi_{0}^{}=2\Gamma_{0\to1}^{(\rm R)}\Gamma_{2\to1}^{(\rm R)}$ and 
\begin{eqnarray}
\Lambda_p &=&\phi_{0}^{}+p\,
\left(\Gamma_{2\to1}^{(\rm R)}-2\Gamma_{0\to1}^{(\rm R)}\right)
\Gamma_{1\to1-p}^{(\rm L)}\nonumber\\
{\mathcal D}_0 &=&\Gamma_{1\to0}^{(\rm
   L)}\Gamma_{2\to1}^{(\rm R)}+2\Gamma_{0\to1}^{(\rm
   R)}\left(\Gamma_{2\to1}^{(\rm
     R)}+\Gamma_{1\to2}^{(\rm L)}\right)\,.
\end{eqnarray}
Expression (\ref{eq:conductance}) can change sign depending on the factor
$\Lambda_{p}$. For $G$ along
$\left|n+1,1\right>\to\left|n,2,l_{\rho},l_{\sigma}\right>$ we have $p=1$ ,
and $p=-1$ for $G$ along
$\left|n+1,1\right>\to\left|n,0,l_{\rho},l_{\sigma}\right>$.  Consider the
transition $\left|n+1,1\right>\to\left|n,2\right>$, thus $p=+1$.  Along this
line, $\Gamma_{2\to1}^{(\rm R)}$ has contributions from the excited states,
while the other rates are $\Gamma_{0}^{(\rm L,R)}$. For
non-interacting leads, $g_{0}=1$, the condition $\Lambda_{1}\leq 0$ is
equivalent to
\begin{equation*}
\left[\frac{\Gamma_{0}^{(\rm R)}}{\Gamma_{2\to 1}^{(\rm R)}}-
\frac{1}{2}\right]A\geq 1\, .
\end{equation*}

We define $A_{\rm c}$ as the critical asymmetry above which NDC is found. In
region I, $A_{\rm c}^{({\rm} I)}=2$. In region II, $A_{\rm c}^{({\rm
    II})}=2\left(2g_{\sigma}^{}+1\right)/\left(2g_{\sigma}^{}-1\right) >
A_{\rm c}^{({\rm I})}$. Thus, for $A>A_{\rm c}^{({\rm II})}$ one has NDC in
both regions I and II.  On the other hand, the condition for having NDC in
region III is
\begin{equation}
\frac{A}{2}\left(\frac{2g_{\rho}^{}
g_{\sigma}^{}-g_{\rho}^{}-g_{\sigma}^{}}{2g_{\rho}^{}
g_{\sigma}^{}+g_{\rho}^{}+g_{\sigma}^{}}\right)\geq 1\,.
\label{eq:condition1}
\end{equation}
For small exchange interactions this requirement cannot be fulfilled.  Indeed
for $V_{\rm ex}\le V_0/2$ one has
$g_{\sigma}^{2}+3g_{\rho}^{2}-4g_{\rho}^{2}g_{\sigma}^{2} > 0$ which cannot be
satisfied simultaneously with (\ref{eq:condition1}). We conclude that for
sufficiently large asymmetry NDC is generally found in regions I and II and
PDC in region III at the transition $\left|n+1,1\right> \to\left|n,2\right>$.

The conditions for NDC along the lines
$\left|n+1,1\right>\to\left|n,0,0,1\right>$ and
$\left|n+1,1\right>\to\left|n,0,1,0\right>$ is
\begin{equation*}
\left[\frac{1}{2}-\frac{\Gamma_{0}^{(\rm R)}}
{\Gamma_{2\to 1}^{(\rm R)}}\right]A\geq 1\, .
\end{equation*}
It is clear that both in I and in II no NDC are found. In zone
III, however, it is possible to have $\Gamma_{2\to1}^{(\rm R)} >
2\Gamma_{0}^{(\rm R)}$ with NDC for 
\begin{equation*}
A > A_{\rm c}^{({\rm III})}=2\,
\frac{g_{\rho}+g_{\sigma}+2g_{\rho}g_{\sigma}}
{g_{\rho}+g_{\sigma}-2g_{\rho}g_{\sigma}}\,
\end{equation*}
If $A > \max\{A_{\rm c}^{({\rm I})},A_{\rm c}^{({\rm II)}},A_{\rm c}^{({\rm
    III})}\}$ the NDC-PDC pattern has therefore a
''photographic-negative''-like shape. This means that if the line
$\left|n+1,1\right>\to\left|n,2\right>$ ($p=+1$) has NDC, the two adjacent
ones ($p=-1$) have PDC, and {\em vice versa}. This feature can be
found in many regions of the density plot for not too high bias voltage (cf.
Fig.~\ref{fig:manyn}).

These results indicate that in addition to the condition
$\varepsilon_{\rho}\neq \varepsilon_{\sigma}$ one must have non-integer steps
in the transition rates as a function of the energy in order to find NDC. It
seems that this is a genuine feature in present case of an interacting system
with non-Fermi liquid correlations, i.e. with spin-charge separation.

\subsection{Spin-flip relaxation}
\label{sec:phasediag}
Since the NDC is related to the occupation of states with spins higher than
that of the ground state, it is of great interest to introduce into the master
equation an extra spin-flip relaxation rate $\Gamma_{s\to s'}^{w}$. This will
give insight into the depletion behavior of the NDC when spin-flip processes
are present. We assume
\begin{equation}
\Gamma_{s\to s'}^{w}=\begin{cases}
w&|s'|<|s|\\
w \exp\left[-\frac{1}{2}\beta E_{\sigma}^{}(s'^2-s^2)\right]&|s'|>|s|
\end{cases}\, ,
\end{equation}
with $s'=s\pm 2$, $n' = n$, and consider $w$ to be an
estimate of the typical spin-flip rate. This spin-flip mechanism
acts against the trapping of the higher spin states, providing an escape
possibility $\left|n,2\right>\to\left|n,0\right>$. This is expected to
decrease the chance to have NDC.

The differential conductance for different values of the spin relaxation rate
$w$ is shown in Fig.~\ref{fig:ndc22} for voltages within the five-states
region, $s_{\rm max}=\pm 2$. The peaks for a positive bias are reduced by
increasing $w$, especially the NDC state which eventually becomes positive.
For negative bias, electrons traverse a lower barrier tunneling into the dot,
and a higher barrier tunneling outside: $P_{n,0},P_{n,\pm 2} < P_{n+1,\pm 1}$.
Therefore the relaxation from the states $\left|n,\pm 2\right>$ is almost
ininfluent, since these states have already a small occupation probability.
\begin{figure}[htbp]
  \begin{center}
\includegraphics[width=7cm,keepaspectratio]{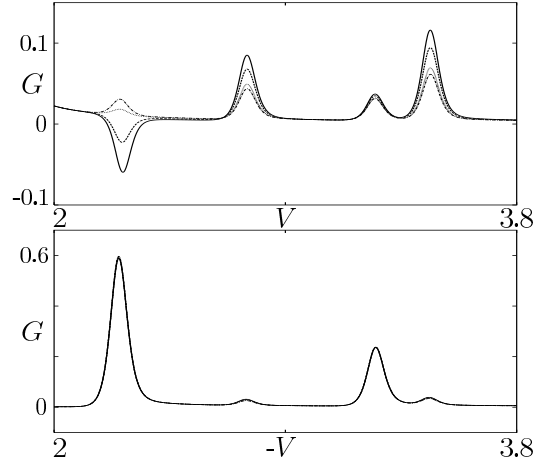}
    \caption{Differential
      conductance (units $e^2\Gamma^{\rm R}_0/E_{\sigma}$) a function of
      positive and negative voltages $V$ (units $E_{\sigma}/e$) in the
      five-states region for $n_{\rm g}=0.485$ for spin
      relaxation rates $w=0.2$ (solid), 0.3 (dashed), 0.5 (dotted) and 0.6
      (dash-dotted) in units $\Gamma_{0}^{(\rm L)}(E_{\sigma}^{}/\omega_{\rm
        c}^{})^{\alpha}_{}$, for $k_{\rm B}T=10^{-2}E_{\sigma}$,
$E_{\rho} = 25 E_{\sigma}$, $A=10$,
      $\eta=1/2$, $g_{0}=0.9$, $g_{\rho}=0.7$ and $g_{\sigma}=1.25$.}
    \label{fig:ndc22}
  \end{center}
\end{figure}

As an example, we consider the above five-states model including spin-flip
relaxation. The conductance has the same form as in (\ref{eq:conductance})
\begin{equation}
G=\frac{e\phi_{w}^{}}{{\mathcal D}_{w}^2}\sum_{p=\pm
    1}^{}\Lambda_{p}^{w}\partial_V\Gamma_{1\to1+p}^{(\rm L)}(V)
\end{equation}
but now $\phi_w=2\Gamma^{(\rm R)}_{0\to1}(w+\Gamma_{2\to1}^{(\rm R)})$
and 
\begin{eqnarray}
\Lambda_{p}^{w}&=&\phi_w+p\left[\Gamma_{2\to1}^{(\rm
    R)}-2\Gamma_{0\to1}^{(\rm R)}\right]\Gamma_{1\to1-p}^{(\rm L)}\nonumber\\
{\mathcal D}_w &=& \Gamma_{1\to 0}^{(\rm L)}\Gamma_{2\to
  1}^{(\rm R)}+2\Gamma_{0\to 1}^{(\rm R)}\left[\Gamma_{2\to 1}^{(\rm
    R)}+\Gamma_{1\to 2}^{(\rm L)}\right]\nonumber\\
&&+w\left[\Gamma_{1\to 0}^{(\rm L)}+\Gamma_{1\to 2}^{(\rm L)}+
2\Gamma_{0\to  1}^{(\rm R)}\right]\, .
\end{eqnarray}
As above, the sign of the conductance is solely given by ${\Lambda}_{p}^{w}$
which is now a function both of the asymmetry and of the relaxation ${w}$. We concentrate on the region I ($2E_{\sigma}< eV <
\varepsilon_{\sigma}$) of the line $\left|n+1,1\right>\to\left|n,2\right>$.
The condition ${\Lambda}_{1}^{w}\leq0$ reduces at $T=0$ to
\begin{equation}
\label{eq:phase}
\frac{w}{\Gamma_{0}^{(\rm
        L)}}\le\left(\frac{2E_{\sigma}^{}}
{\omega_{\rm c}^{}}\right)^{\alpha}_{}
\left[1-\frac{1}{2}\left(\frac{eV}{eV-2E_{\sigma}^{}}\right)^{\alpha}_{}
-\frac{1}{A}\left(\frac{eV}{2E_{\sigma}^{}}\right)^{\alpha}\right]
\nonumber\\
\end{equation}
with $\alpha$ given in (\ref{eq:alpha}).
\begin{figure}[htbp]
  \begin{center}
    \includegraphics[width=7cm,keepaspectratio]{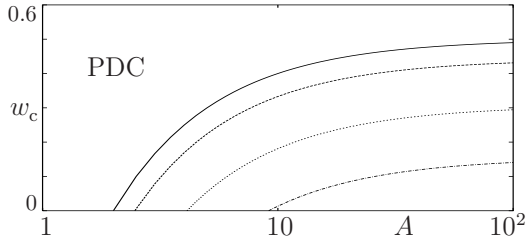}
    \caption{Phase diagram for the critical value of
      the spin-flip relaxation rate $w$ (units $\Gamma_{0}^{(\rm
        L)}(E_{\sigma}^{}/\omega_{\rm c}^{})^{\alpha}_{}$) in region I as a
      function of the asymmetry $A$ for $T=0$ and different interaction
      parameters in the leads: $g_{0} = 1.0$ (solid), $g_{0} = 0.9$ (dashed),
      $g_{0} = 0.75$ (dotted) and $g_{0} = 0.65$ (dash-dotted).}
    \label{fig:ndc2}
  \end{center}
\end{figure}
The critical value $w_{\rm c}$ for the crossover is plotted in
Fig.~\ref{fig:ndc2} as a function of $A$. At a given asymmetry, for $w>w_{\rm
  c}^{}$ the conductance peak crosses over to positive values. The stronger
the interaction in the leads the smaller is the threshold in the relaxation
for destroying the NDC peak.

\section{Conclusions}
\label{concl}
We have investigated the non-linear current voltage characteristic of a 1D
quantum dot described by a Luttinger liquid of a finite length connected
via tunnel barriers to interacting quantum wires. The system contains
four characteristic energy scales: charge and spin addition energies, and the
energies needed to excite charge and spin density collective modes in the
quantum dot. We have discussed in particular the effects related to the
presence of the electron spin.

In the linear regime, we have reproduced the parity effect in the distances
between the Coulomb blockade conductance peaks.  In contrast to the case
without interaction discussed earlier,\cite{baranger} the temperature behavior
of the peak positions shows signatures of the non-Fermi liquid nature of the
leads. At temperatures higher that the spin addition energy, the peaks become
equidistant.

In non-linear transport, we have found NDC features that are related to the
presence of spin-charge separation. They are connected to states with higher
spins that participate in the electron transport.

This effect is different from the spin blockade phenomenon discussed
previously.\cite{w95} In the spin blockade of type I, a ``trapping'' argument
is used to explain the effect. However, in this case the only state that leads
to trapping (and thus can decrease the current) is a state with the highest
total spin and highest energy for a given electron number. Thus, only one NDC
peak can be present in a given $n \leftrightarrow n+1$ transition. The
``trapped'' status is achieved because of the Clebsch-Gordan coefficients that
are introduced in the calculation of the transition rates.

In contrast, in the present case, each state with even (odd) spin can become
trapped if $(A-1)V > 0$ ($< 0$) such that the transition lines that populate
such a state exhibit NDC. Whether or not such a state is eventually
trapped is determined by the transition rates which are here microscopically
evaluated using the Luttinger liquid model. The result is that even if the
system contains asymmetric barriers that lead to trapping of the appropriate
candidate states in the dot, it is still necessary to have charge-spin
separation in order to obtain NDC. It is clear from the above that the effect
of spin-charge separation is not only the trivial removal of energetic
degeneracy of the CDW and SDW states. The spectral weights
contained in the tunneling rates (\ref{eq:ratesum}) play a crucial role in
determining whether or not a particular transition line corresponds to NDC.

Since NDC is driven by asymmetry, there must be a threshold value $A_{\rm c}$
above which the phenomenon occurs. This critical value is not universal. Each
transition line has a different critical value for the asymmetry.

We have assumed that the collective excitations that do not alter the
$z$-component of the spin in the quantum dot have infinitely short relaxation
times as compared with the excitations which are associated with spin changes.
We feel that this assumption is justified in view of recent results suggesting
that in quantum dots the relaxation of states without flipping spins can be
orders of magnitudes shorter than that associated with spin
flips.\cite{nazarov2} Nevertheless, the question whether or not without this
assumption the predicted NDC-phenomena would disappear is legitimate.
In order to answer this question quantitatively, it
is necessary to repeat the calculations including the collective
states of the quantum dot as dynamical variables in the master
equation.\cite{kinaret}

Spin-flip relaxation processes only seem to weaken existing NDCs, as indicated
by the results in the section \ref{sec:phasediag}.  Thus, we can expect that
by removing relaxation processes and including additional stable states in the
transport process without flipping the total spin will not change
qualitatively the trapping mechanism described in the section \ref{pdcndc}.
The NDC predicted here, which appears to be a consequence of the trapping
mechanism and the separation of energy scales for spin and charge excitations
together with the Luttinger liquid features that enter the rates, will not be
depleted.

In experiment, several possibilities for measuring the predicted non-linear
phenomena exist. The necessary condition for applying the above model is
\begin{equation}
  \label{eq:condition}
  k_{\rm B}T\ll \varepsilon_{0}\ll E_{\rm F}\,,
\end{equation}
with $\varepsilon_{0}=\pi v_{\rm  F}/a$. This is equivalent to 
\begin{equation}
  \label{eq:lengths}
  \lambda_{T}\gg a \gg \lambda_{\rm F}
\end{equation}
where $\lambda_{T}=\pi v_{\rm F}/k_{\rm B}T$ is the thermal length and
$\lambda_{\rm F}=2\pi/k_{\rm F}$ the Fermi wave length. Tunnel barriers will
in general be asymmetric in any case.

In semiconductor based quantum wires, quantum dots are prepared via depleting
the electron density such that eventually an electronic island is formed
accidentally between two impurities.\cite{a00} In this system, the Fermi
energy appears to be relatively low, $E_{\rm F}\approx 2$\,meV, due to the
almost depletion of the lowest sub-band.\cite{k00} The dot energy level
spacing is about 1\,meV for a quantum dot of length $a\approx 0.2\,\mu$m and
an effective mass $m^{*}\approx 0.07\,m_{\rm e}$ (for GaAs). These parameters
at the first glance seem to be outside the above region of validity of the
model, although the condition with respect to the temperature is easily
fulfilled. For a quantum dot of larger size ($a\approx 1\,\mu$m) the situation
would be much more favorable. This seems to be achievable in other
semiconductor based quantum dot systems, such as those fabricated by scratching
techniques.\cite{haug} With this one could fabricate an appropriately scaled
quantum dot in a quantum wire. Parameters to be achieved should be: width of
wire $\approx 50$\,nm in order to make a Fermi energy of a few meV achievable;
distance between the tunnel barriers (point contacts) that define the quantum
dot $\approx 1$\,$\mu$m giving $\varepsilon_{0}\approx 0.1$\,meV. Temperatures
should be well below 1K. With this technique, one would have the advantage of
being able to adjust the asymmetry of the tunnel barriers.

In carbon nanotubes the situation with respect to the energy scales seems to
be even more favorable since $E_{\rm F}\approx 2$\,eV, $v_{\rm F}\approx
8\cdot10^5$m/s and the level spacing $\epsilon_{0}\approx 5$\,meV with
temperatures of the order $T\approx 100$mK.\cite{cetal98} However, here the
non-interacting energy spectrum consists of four branches including the spin.
The present theory has to be adjusted to this case in order to apply the
results.  Again, due to the fact that the predicted phenomenon seems to be
quite generally valid, one can expect that experiments on carbon nanotubes
should show NDC associated with the higher spin states.

\section*{Acknowledgments} 
We are grateful for helpful and illuminating discussions with Rolf
Haug. Financial support by the European Union via
TMR-networks FMRX-CT98-0180 and HPRN-CT2000-0144, and from the Italian
MURST PRIN02 is gratefully acknowledged.

\end{document}